\documentclass[journal]{vgtc}                

\ifpdf
  \pdfoutput=1\relax                   
  \usepackage{graphicx}                
  \DeclareGraphicsExtensions{.pdf,.png,.jpg,.jpeg} 
\else
  \usepackage{graphicx}                
  \DeclareGraphicsExtensions{.eps}     
\fi%

\graphicspath{{figures/}{pictures/}{images/}{./}} 

\usepackage{microtype}                 
\PassOptionsToPackage{warn}{textcomp}  
\usepackage{textcomp}                  
\usepackage{mathptmx}                  
\usepackage{times}                     
\usepackage{cite}                      
\usepackage{tabu}                      
\usepackage{booktabs}                  
\usepackage[colorinlistoftodos,prependcaption,textsize=tiny]{todonotes}

\definecolor{emerald}{rgb}{0.31, 0.78, 0.47}

\definecolor{britishracinggreen}{rgb}{0.0, 0.26, 0.15}

\definecolor{mygrey}{rgb}{0.5, 0.5, 0.5}

\definecolor{revisioncolor}{rgb}{0.97, 0.61, 0.48}
\newcommand\rev[1]{{\color{black}{#1}}}

\onlineid{1239}

\vgtccategory{Research}
\vgtcpapertype{Application/Design Study}

\title{Visilant: Visual Support for the Exploration and Analytical Process Tracking in Criminal Investigations}

\author{Krist\'{i}na Z\'{a}kop\v{c}anov\'{a}, Marko \v{R}eh\'{a}\v{c}ek, Jozef B\'{a}trna, Daniel Plakinger, Sergej Stoppel, Barbora Kozl\'{i}kov\'{a}}

\authorfooter{
\item
 K. Z\'{a}kop\v{c}anov\'{a}, M. \v{R}eh\'{a}\v{c}ek, J. B\'{a}trna, D. Plakinger, and B. Kozl\'{i}kov\'{a} are with the Masaryk University. E-mails: zakopcanova.k@gmail.com, rehacek@mail.muni.cz, 445668@mail.muni.cz, daniel.plakinger@mail.muni.cz, kozlikova@fi.muni.cz.
 \item S. Stoppel is with the University of Bergen and Rainfall AS, Bergen. E-mail: sergejsto@googlemail.com.
}

\shortauthortitle{Z\'{a}kop\v{c}anov\'{a} \MakeLowercase{\textit{et al.}}: Visilant}

\abstract{The daily routine of criminal investigators consists of a thorough analysis of highly complex and heterogeneous data of crime cases. Such data can consist of case descriptions, testimonies, criminal networks, spatial and temporal information, and virtually any other data that is relevant for the case. Criminal investigators work under heavy time pressure to analyze the data for relationships, propose and verify several hypotheses, and derive conclusions, while the data can be incomplete or inconsistent and is changed and updated throughout the investigation, as new findings are added to the case. Based on a four-year intense collaboration with criminalists, we present a conceptual design for a visual tool supporting the investigation workflow and Visilant, a web-based tool for the exploration and analysis of criminal data guided by the proposed design. Visilant \rev{aims to support namely the exploratory part of the investigation pipeline}, from case overview, through exploration and hypothesis generation, to the case presentation. Visilant tracks the reasoning process and as the data is changing, it informs investigators which hypotheses are affected by the data change and should be revised. \rev{The tool was evaluated by senior criminology experts within two sessions and their feedback is summarized in the paper. Additional supplementary material contains the technical details and exemplary case study.}%
} 


\CCScatlist{ 
 \CCScat{K.6.1 TODO}{Management of Computing and Information Systems}%
{Project and People Management}{Life Cycle};
 \CCScat{K.7.m TODO}{The Computing Profession}{Miscellaneous}{Ethics}
}

\teaser{
  \centering
  \includegraphics[width=\linewidth]{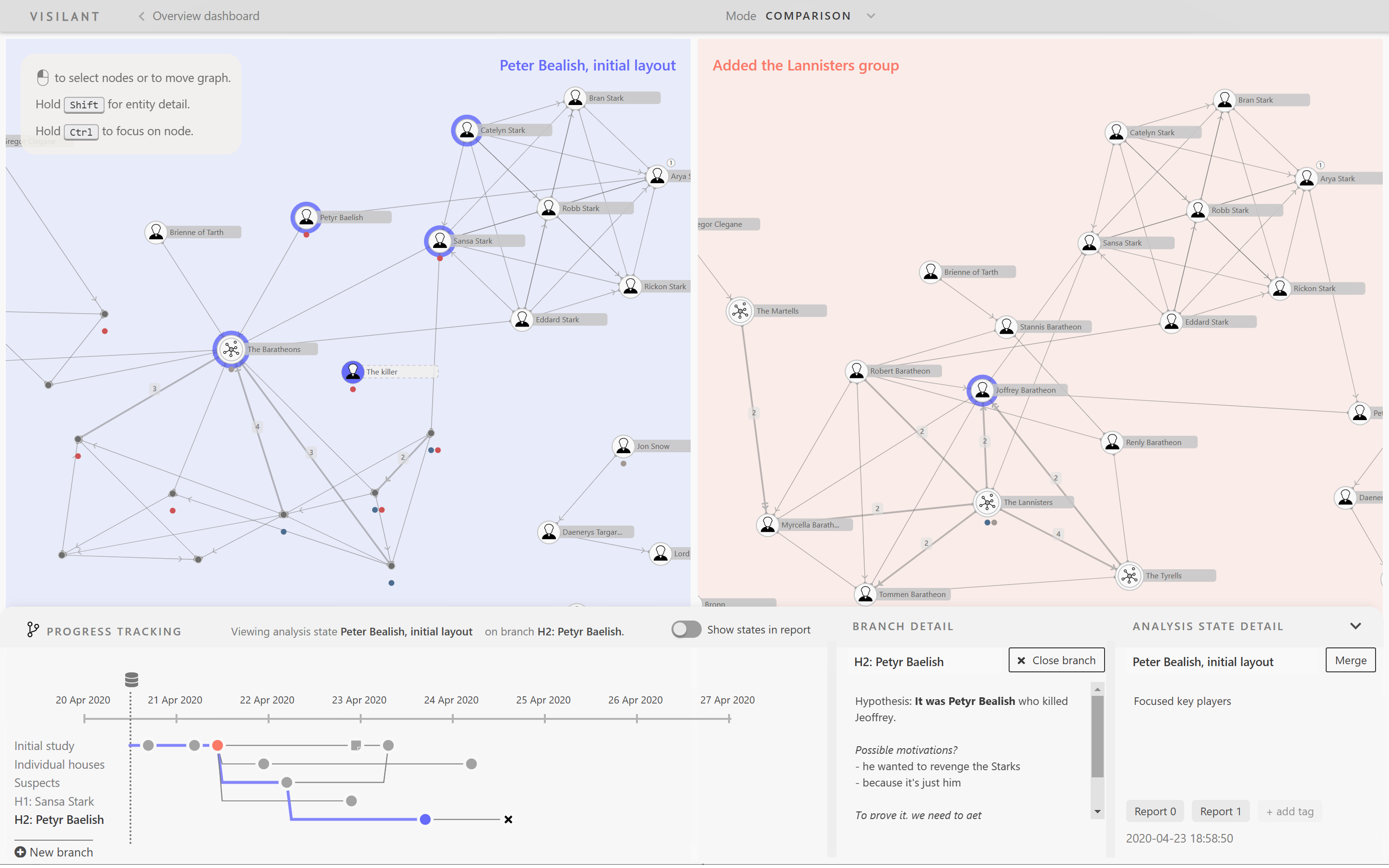}
  \caption{Overview of the core parts of the proposed Visilant tool. The top part shows two states within the investigation process that need to be compared, the bottom part enables users to track the investigation progress, trace it back, and perform required operations.}
	\label{fig:teaser}
}

\vgtcinsertpkg

\begin{document}


\firstsection{Introduction}

\maketitle
Criminology is a multidisciplinary field with roots already in the 18th century when the main aim was to reform criminal law and ground it in society. The term criminology was finally coined in the late 19th century and since then, criminology extended and shifted mostly to the investigation of crimes and their causes. The ingenuity of criminals goes in hand with the necessity to perform vast and complex investigations, develop several hypotheses, and, in the end, solve the crime.
Nowadays, in the digital era, criminal investigators are facing new challenges, coming namely from the necessity to explore extremely large amounts of data related to each case. The new technologies also allow a better organization, communication, and interconnection of criminal networks. Such data is highly heterogeneous, spanning from financial transfers, recorded communication, or photographs from the surveillance systems, to the content of hard drives captured within home inspections. Since criminals are always trying to hide their fraudulent intentions, the exploration of such data is far from being straightforward, as the network of suspicious subjects is usually vast, with many straw men hiding the criminal actions. Therefore, the investigation is a highly cognitively demanding process of thorough exploration and analysis of information with a variety of tools, specifically designed for individual sub-tasks.
Another challenge comes from the time constraints posed to the investigation -- it spans from a couple of days, requiring very fast reasoning, to months or even years for severe fraudulent cases with large and complex networks of suspects.  

The exploration process of a given case is complex and non-linear and investigators need to analyze a multitude of possible scenarios. Thus, their workflow needs to be adequately supported to help them externalize thinking and reasoning. Such aid should reflect their mental model in a simple and comprehensible way and most importantly to interact with it accordingly to support the change of the visual representation of the model with respect to the current state of the investigation. This helps not only in their exploration process within the investigation made by individual criminalists, but also in sharing the knowledge with co-investigators, using their pieces of a puzzle to build the complete model of the case, and, in the end, to prepare the materials as a part of the evidence for the state attorneys. 

The analytical workflow could be divided into three phases. The first phase covers the initial data gathering and storage in a central database, the second comes the data exploration phase, and the last phase consists of developing inferences and conclusions and assembling the final report as a basis for bringing criminal charges. The first phase interleaves with the second one, as it is practically impossible to obtain all necessary data and facts already before the investigation -- as the investigation proceeds, the database is constantly updated with new findings. Based on the initial data, investigators form preliminary inferences and hypotheses that need to be confirmed or rejected. For this process, the appropriate visual representation and means for its exploration are crucial. 

As the most \rev{prominent and important} information in the data is the relations between different objects, the most \rev{commonly} used visual representation \rev{by criminalists} is the network graph. \rev{This was revealed within numerous discussions with the experts and observations of their daily routines. } 
For the investigation process, it is \rev{necessary} that the network analysis supports incorporating new data, incoming throughout the investigation, and allows users to manually insert additional information or assumptions directly using the visual interface and most importantly, visually distinguishes these from the data loaded from the central database.

To ease the cognitive load imposed on investigators, the visual representation should \rev{enable them to carry} as much information and knowledge about the whole investigation as possible. \rev{At the same time, it needs to be carefully designed with respect to the amount of displayed information that can easily lead to visual clutter}. Therefore, in our solution, we propose a specific visual representation keeping the provenance of the investigation, which enables criminalists among other things to trace back the investigation scenarios by exploring their individual branches, to cut false directions, and to merge analysis states.
Our long-term collaboration with criminalists led us first to the creation of the concept for designing the visual support for the field of criminal investigations. It was iteratively derived from numerous discussions and interactive sessions with the chief criminalists from the National Center for Combating Organized Crime (NCCOC) of the Czech Republic. To demonstrate the application of this concept in action, we introduce Visilant, a web-based application supporting the visual exploration of heterogeneous data in criminal investigations. The central part of Visilant consists of the Network analysis view, enhanced by many visual and interaction clues. Experts are also provided with a tailored Progress tracking diagram for the exploration of the history and current progress of the investigation. \rev{The Progress tracking diagram allows the inspection of important analysis states, which are organized into branches representing individual investigation scenarios.} Visilant also supports a fast and intuitive generation of the report for state attorneys, extracted directly from the proposed visual representations.

To summarize, the main contributions of our paper are the following:
\begin{itemize}
\item The conceptual design for visual representations supporting \rev{externalization, progress tracking, and collaboration in} criminal investigations, addressing the domain-specific requirements. 
\item Visual support for non-linear data exploration, reflecting the nature of the investigation process as criminalists need to explore various hypotheses, which often lead to a complex network of investigation branches. 
\item Visilant, a web-based tool for the visual exploration of \rev{relationships in} data analyzed by criminalists, supporting the crucial tasks performed within the investigation process.
\end{itemize}

\section{Background}
\label{sec:background}
In this section, we provide readers with details of the criminal investigation process, its pitfalls, challenges, and the terminology used. 
As each investigated case relates to a different type of crime and operation, it also requires different scenarios of investigation and can be carried out in different time spans.
Some investigations can spread over several months and within that time, criminalists need to explore many different scenarios and aspects of the crime.
On the other hand, there are situations when investigators must act very quickly and come up with inferences and conclusions in a couple of hours. 
In both situations, investigators need to be capable of analyzing individual aspects of the crime in detail, while at the same time keeping in mind the overall “big picture” of the case. This is \rev{important} for valid decision-making but poses significant demands on their mental capacity and memory. 

The quality of captured data has a significant impact on the outcome of the investigation. The most common and severe problems are incomplete, incorrect, and inconsistent data, which can lead to false conclusions. Therefore, the investigator needs to be informed about these facts and keep that in mind within the whole exploration process. 
Xu and Chen~\cite{Xu:2005:CriminalNetwork} formulated these three categories of false data in the following way. \textit{Incompleteness} in data mostly comes from the fact that criminals are trying to minimize mutual interactions or obfuscate their direct connections by using an intermediate straw man in communication. As a result, investigators obtain an incomplete network of communication and relationships and they need to reconstruct the missing nodes and links within their investigation. The second, very common problem, is data \textit{incorrectness}. Investigators often face incorrect data regarding criminals' identities, characteristics, addresses, etc., originating in unintentional data entry errors or intentional deceptions by criminals.  The third problem, data \textit{inconsistency}, occurs when a criminal is investigated by multiple criminalists who are entering the data to the database. \rev{As a consequence}, the database can contain several\rev{, not necessarily consistent,} records of the same person and can thus lead to false assumptions that a single criminal appears to be different individuals. All these problems make the data exploration more complicated and need to be addressed in the visual exploration process.

The visual representations and interaction options proposed in our solution are part of a long-term research project, whose ultimate goal is to provide criminalists with a unified and solid framework for the investigation process, covering their whole workflow. The proposed framework integrates algorithms, methods, and technologies for efficient data storage, mining, annotation, analysis, exploration, and extraction. Among the most prominent features belong the advanced database searches, the similarity-based searches performed on a variety of data (text, images), and the support for their visual exploration. This platform is the result of a four-year intense collaboration with chief criminalists from the NCCOC of the Czech Republic.
As we are fully aware of the fact that criminalists use dozens of specialized tools and applications developed within the last decades, the ultimate goal is not to fully replace all of them but to create a robust framework covering the fundamental parts of the workflow. In practice, the framework can operate with a multitude of formats in order to be still able to utilize some of the specialized tools. 

To distinguish between the analysis performed using the framework covering the whole investigator’s workflow and the visual analysis part, which forms the main scope of this paper, we introduce the following notation. As \textit{analytical workflow} we define the whole process spanning from data gathering, storage, and exploration, to the generation of reports for the state attorney. The \textit{exploration process} covers the part where investigators utilize the visual representations to understand the relationships in data and by further interaction they can proceed with the investigation and derive conclusions. The proposed visual solution for the exploration process is covered in this paper.  

\section{Related Work}

Several visualization tools for various subtasks of criminal investigations have been introduced over the last decades. In this section, we outline the related work for criminal investigations with a focus on criminal networks and design considerations for investigation systems for criminal intelligence.
For a comprehensive discussion of relevant theoretical frameworks of network analysis in security domains we refer the reader to the collaborative book edited by Masys~\cite{Masys:2014:Networks}, in particular the chapter by Strang~\cite{Strang:2014:NetworkAnalysis}, for the design considerations of network analysis tools.
A review of available tools for representation and externalization in criminal investigations is provided by Passmore et al.~\cite{Passmore:2015:ExternalThinking}.
Xu and Chen~\cite{Xu:2005:CriminalNetwork} discussed the challenges of terrorist networks and existing solutions, concluding that visualization tools revealing structures and interactions within criminal networks are needed. 
Islam et al.~\cite{Islam:2016:TowardsAnalyticalProvenance} outlined requirements for a criminal intelligence analysis system and proposed a technique to capture visual analytical states and processes, known as analytic provenance. We used this work as guidance for the design of the Progress tracking diagram in our system.
Wong et al.~\cite{WilliamWong:2018:Variability} argued that criminal investigations rely on the creative generation of plausible explanations or "storytelling" and presented an interaction design supporting criminal investigators in the storytelling process. In our work, we intrinsically support this creative process through a non-linear exploration of data and hypothesis generation.
Lee et al.~\cite{Lee:2015:DataToStory} took a closer look at how the visualization community uses the storytelling in the proposed tools and conceptualized the required steps to transform data into stories. Our proposed Visilant tool supports all steps of the storytelling process, resulting in a visual report for state attorneys.

The advancements in communication and computing technology introduced a new form of criminal activity, cybercrime. The field of cybersecurity can be to some extent seen as similar to the classical criminal investigation. While cyber criminology introduces its own challenges, such as the immense size of networks of bots or coordinated attacks, investigations of cybercrimes are often interleaved with classical criminal investigations and require the usage of similar tools. Therefore, we briefly point out a few related works focusing on the network visualization of cybercrime data.
Glanfield et al.~\cite{Glanfield:2009:OverFlow} argued that only a small portion of network data can be inspected at a time and proposed a network representation providing context for alternate visualizations by displaying a high-level view of network events. 
Tsigkas et al.~\cite{Tsigkas:2012:VisualSpam} focused on spam-bot networks and provided security analysts with a tool for interactive reasoning about large scale spam attacks. 
Chu et al.~\cite{Chu:2010:AttackGraphs} presented NAVIGATOR, a visualization tool for attack graphs depicting the attacked network topology and device infrastructure.
Erbacher~\cite{Erbacher:2012:CCGC} introduced a novel high-level visual design of network states that allows users to review the network state and locate potentially problematic anomalies.
Wong et al.~\cite{Wong:2016:HAT1} and Gronewald et al.~\cite{Groenewald:2017:HAT3, Groenewald:2017:HAT2} summarized their findings on the process of sense-making in criminal intelligence analysis in a series of publications. We considered their findings when designing our tool.

The scientific community early realized that the exploration processes benefit from provenance information and the ability to return to and inspect previous exploration states. 
Ragan et al.~\cite{Ragan:2016:ProvenanceCharact} characterized different types of provenance information and purposes and discussed why they are desired in visual analytics.
VisTrails, presented by Callahan et al.~\cite{Callahan:2006:VisTrails}, displays workflow evolution provenance as a history tree, where each node represents a workflow that generates a visualization. 
Focusing on data mining frameworks, Kreusler et al.~\cite{Kreuseler:2004:History} introduced history management that allowed the reordering of history states.
Heer et al.~\cite{Heer:2008:GraphicalHistories} presented a graphical history tool for visual analysis support, storing visualization states, and allowing users to navigate between them.
Shrinivasan et al.~\cite{Shrinivasan:2008:Reasoning} developed an information visualization framework consisting of three main view components, data view, navigation view, and knowledge view, that allows users to revisit and evaluate previous visualization states.
Stitz et al.~\cite{Stitz:2018:KnowledgePearls} presented KnowledgePearls that structured the visualization states as provenance graphs and allowed for querying and exploration of the visualization states based on their state similarity.
Gratzl et al.~\cite{Gratzl:2016:FromExploration} introduced a visualization model that integrates data exploration and presentation of discoveries, allowing for transitions between exploration and presentation states.
To tackle the analysis of large multivariate graphs, Nobre et al.~\cite{Nobre:2018:Juniper} proposed Juniper, a visualization technique that linearly lays out the tree and juxtaposes the nodes with a table visualization.
Mathisen et al.~\cite{Mathisen:2019:InsideInsights} proposed InsideInsights, a web-based system for network state and structure annotation, designed for a dynamic collaborative work environment.
Focusing on an unstructured dynamic investigation process, our solution also provides users with the provenance information of the investigation process in the form of a non-linear tracking diagram.

The exploration of the criminal network is one of the most important tasks of criminal investigations. As such, a fitting visual representation of the network is a necessity. In the following, we briefly outline some of the related works focusing on network visualization. We refer to a chapter by Kerren et al.~\cite{Kerren:2014:IntroNetworkVis} for a comprehensive introduction to Multivariate Network Visualization. Furthermore, we acknowledge the comprehensive state-of-the-art report on group structures in graphs provided by Vehlow et al.~\cite{Vehlow:2015:TheSO}, as well as the state-of-the-art report on multivariate graph visualization by Nobre et al.~\cite{Nobre:2019:MultivariateNetworksSTAR}.
Lee et al.~\cite{Lee:2006:GraphTasks} proposed a taxonomy of low-level tasks for graph visualization that can be combined into advanced graph functions.
Van den Elzen and van Wijk~\cite{Elzen:2014:MultiNetwork} focused on non-expert users and proposed a visualization tool for multivariate network exploration that allows them to proceed from detail to overview via selections and aggregations, producing high-level infographic-style overviews. As networks increase in size and they become cluttered and harder to comprehend, aggregations are popular technique to deal with large network data. Such an approach was also taken by Shi et al.~\cite{Shi:2014:HierarchicalNetwork} with OnionGraph that allows node aggregation based on node attributes or topology. 

\rev{Extensive and complex analyses of criminal investigation data impose significant demands on criminalists' cognitive capacity and thus need to be considered.  
Sweller~\cite{low:2005:modality} describes the cognitive load as the relationship between the working memory and cognitive demands of a particular task. Chandler et al.~\cite{chandler:1991:cognitive} distinguish three components of the cognitive load: the intrinsic, germane, and extraneous load, each affecting learning and decision making. Therefore, the visualization tool design decisions effectively influence the amount of load placed on the working memory and thus the executive attention and performance~\cite{engle:2002:working, seufert:2007:impact}.
Huang et al.~\cite{Tony2009} analyzed the effectiveness of network representations with respect to cognitive load.}
\rev{Other influential works deal with the support of collaborative environments of network visualizations, such as work of Mahyar and Tory~\cite{Mahyar:2014}, proposing visual support for externalization and merging network diagrams of collaborators. 
In our solution, we differ in showing two network layouts side by side before their merge and in the option to decide and manually select parts to be merged.
}

\section{Analysis of Requirements}
\label{sec:requirements}
In tight collaboration with the criminal investigators of the 
NCCOC of the Czech Republic, we iteratively investigated and compiled the requirements for their exploration process. We followed the exploratory and user-driven design process to help us identify the requirements ~\cite{ghani2013visual, sedlmair2012relex, meyer2019criteria}. 

Within the four years of discussions and exploration of criminalist workflow and needs, we went through several stages. In the beginning, we conducted a series of general interviews \rev{with criminalists of different positions} to understand their work process and identify their biggest challenges. These were discussed in larger groups, as well as with individual investigators. At the same time, we studied domain literature to get a general background and prepare ourselves for the next phase. 

In the second phase, we carried out multiple contextual inquiries, where we observed in detail their daily routine. We focused namely on the tools they used, how they interacted with them, and what were their main obstacles and limitations. Except for the observation, we conducted post-observation interviews, where we asked the investigators to name the problems and limitations of their current workflow. Within the interviews, we also discussed positive aspects of their daily routine with a specific focus on the beneficial features of tools in use. 

Based on this input, we started an iterative process \rev{of our system development} that consisted of designing appropriate visualization solutions, creating low fidelity prototypes, and discussing these with the investigators. The iterative process proved to be crucial not only for further understanding of their problems and refining the tasks but also for establishing their understanding that visual analysis can aid them way more than they originally imagined. This sparked their creativity and they started bringing up various situations and ideas where they would appreciate our assistance. In line with this process, we were designing and implementing the prototypes of the proposed visual representations, which served as the main basis for the discussions and our envisioned Visilant tool. \rev{These prototypes were always presented to the investigators and their feedback was incorporated
into the next iterations so that the prototypes were in line with their requirements.}
In the last stage, we implemented the final prototype of our proposed Visilant tool and presented it to the investigators within \rev{two interactive sessions (one remote and one in person)}, described in detail, along with its results, in Chapter~\ref{sec:feedback}.
\rev{As we paid specific attention to discussing the progress and results with the chief criminalists from the NCCOC of the Czech Republic, we believe that it increases the chance of the proposed system being adopted by criminalists on the national level.}

In the rest of this chapter, we summarize the most substantial requirements collected within the first two phases described above. Based on their focus, we divide them into two parts. The first part contains requirements on the exploration process of the criminal investigation and the second part consists of more specific requirements for the network visualization. 

\subsection*{Requirements for the Exploration Process}
    \noindent\textbf{\rev{R1:} Support for non-linear exploration.} The exploration of data within investigation is, by definition, non-sequential, with many side branches and rollbacks. Investigators need to keep track of the individual branches, as well as the big picture of the whole case. 
    
    \vspace{1mm}
    
    \noindent\textbf{\rev{R2:} Comparison and combination of results of individual exploration paths.} Investigators need to formulate multiple hypotheses about the case, which need to be confirmed, denied, or further modified as the investigation proceeds. When a hypothesis is confirmed, it may be important to project the findings to other paths of the exploration on-demand. \rev{This has to be fully controlled by the user and performed on the most detailed level where the user can visually compare the differences between two network states that should be merged.}
    
    \vspace{1mm}
    
    \noindent\textbf{\rev{R3:} Classification of the evidence data.} For the trustworthiness of the evidence, it is \rev{important} to distinguish the sources of evidence and levels of their credibility throughout the visual analysis, to be able to evaluate the data correctly and base their conclusions on them. Such information needs to be easily accessible.
    
    \vspace{1mm}
    
    \noindent\textbf{\rev{R4:} Tracking the impact of newly added data.} Over the course of the investigation, when new evidence and findings are added to the case, these may influence the ongoing investigation. Therefore, investigators need to be well informed about the potentially affected analysis states and related investigation branches and have the option to add new information to the network visual representation.
    
    \vspace{1mm}
    
    \noindent\textbf{\rev{R5:} Non-intrusive collaboration.} Each criminal case is commonly investigated by a team of criminalists. Therefore, an efficient way to share their exploration process and inferences is crucial for the advancement in the investigation. At the same time, the mode of collaboration needs to be designed in such a way that criminalists cannot \rev{anytime} interfere with the analyses performed by their colleagues and influence their work-in-progress. 
    
    \vspace{1mm}
    
    \noindent\textbf{\rev{R6:} Assembling the final report.} Once investigators develop the inferences and conclusions from the investigation, these need to be assembled into a final report, which is then handed in to the state attorney as final evidence for prosecution.

\subsection*{Network Visualization Requirements}
     \noindent\textbf{\rev{R7:} Data manipulation.} As the investigation proceeds, criminalists need to have the ability to modify the existing data and enter new pieces of information. This needs to be performed on several levels, based on the level of information credibility. Criminalists need the option to change or enrich their local network representation with new assumptions or information. When such information is trusted enough, they need a way to share such information with their colleagues.
    
    \vspace{1mm}
    
    \noindent\textbf{\rev{R8:} Missing data indication.} As the investigation data often suffers from incompleteness and inconsistencies, criminalists lack the information about individual nodes and links in the network. However, they still need means to create "placeholder" objects that can be later enriched by newly revealed findings and associated with a real object, e.g., a person. They can also use these placeholder objects to test and observe how the network structure may change with their presence in different scenarios. 
    
    \vspace{1mm}
    
    \noindent\textbf{\rev{R9:} Efficient organization of objects and their relationships.} Criminalists need various ways to introduce structure and semantic meaning into the network visualization. Among those belong the possibility to form semantical groups of objects or annotate them both textually and visually. This goes in line with the requirements for the fundamental interactions with the network, generally following the focus+context and detail-on-demand visualization principles.

\section{Design Concept}
Although the overall process of establishing a collaboration with a certain domain usually follows similar guidelines, in criminal investigations we identified interesting particularities that led us to the description of the domain-specific design process. 
Within our interviews and discussions, we revealed that the most challenging part of the exploration process in investigations is to keep track of the investigation process, as it poses an enormous cognitive load to criminalists. In practice, it means that criminalists explore a multitude of possible scenarios simultaneously and the states of their progress need to be available throughout the whole investigation. These scenarios follow either hypotheses formulated in the initial phase of the investigation or explore questions that appeared throughout the investigation process. Formulating these can lead to the "divide and conquer" investigation approach when individual scenarios can be verified by different criminalists or at different times, and their inferences can be combined at any time.
The final evidence can then be assembled from a set of analysis states, presenting respective inferences.
All this can happen in a matter of days, sometimes even hours, or it can span several months of a tedious investigation. 
Therefore, having visual aids supporting the divergent and convergent phases of the investigation process on their daily work bases is of utmost importance.

To grasp these issues and address them in the visualization design, we present the concept of visual exploration performed via the \textit{visualization documents}, whose states can be captured on demand and accessed in our newly introduced \textit{Progress tracking diagram} (see Section~\ref{sec:tracking}). As the investigation data is of different "quality" based on the presence and credibility of their source, we also present the \textit{credibility levels of data} that define different modes of operation.

\subsection{Visualization Document}
Each investigation can be thought of as a set of investigation scenarios, focusing on distinct hypotheses. Such a scenario can be taken as a structured scheme, starting with an initial subset of data to be explored and then choosing the appropriate visual representation and interactions to fulfill given tasks.
Therefore, we introduce the concept of a visualization document, which is defined by the initial dataset and selection of an appropriate visualization technique. It can be thought of as a parallel to a traditional text document, where we express our thoughts and ideas by adding new lines of text or annotations in the form of comments. In the visualization document, criminalists can interact in an analogical manner with the representation and incorporate their thoughts and findings directly in the visualization, thus enriching the document with new information.
Criminalists could start a new visualization document with every new hunch or hypothesis; however, since these thoughts are related to the same case and use the same data, we could think of them rather as different versions of the same visualization document and provide analysts with a mechanism to store multiple versions and track down the paths of thinking. For that, we introduce the concept of progress tracking, discussed in the following section.

\subsection{Progress Tracking}
\begin{figure}[htb]
  \centering
    \includegraphics[width=1.0\linewidth]{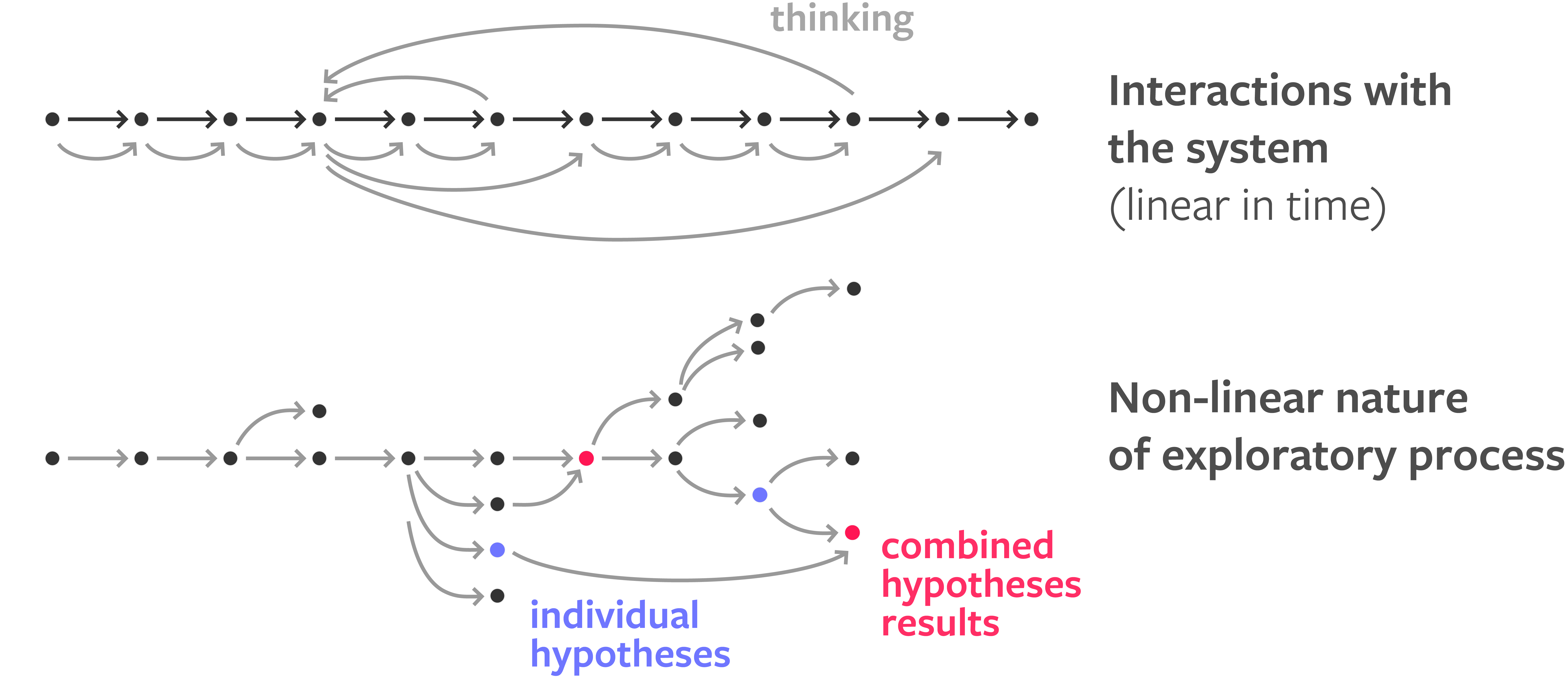}  
\caption{Comparison between the linearity of interactions with a system (top) and the non-linear exploratory process (bottom). The non-linear nature of exploration fits the investigation process by enabling users to create branches, trace their progress and history, terminate them, or combine their content.}
  \label{fig:nonlinear}
\end{figure}

A common approach to tracking the history of a visualization workflow is to use a provenance graph. It tracks users' interactions and provides them with a graph representation storing each interaction as a node. 
However, such graph stores interactions in a linear manner, while our explorative process is by its nature non-linear (see Figure~\ref{fig:nonlinear}). Due to the storage of all actions as individual states, such graphs quickly become very large, thus difficult to navigate in. Even though provenance graphs can offer advanced features, such as a grouping of nodes, these are done based on users' interactions, which do not necessarily mirror their reasoning behind the actions. And so, we intentionally leave it up to the users to define important states to be saved so that it better reflects their mental model.

Therefore, we introduce the concept of progress tracking that aims to \rev{reflect} the whole non-linear investigation process for a particular case \rev{and fulfill a function of an investigator's log book}. 
At any point in the exploration, users can capture their work in the visualization document by creating an \textit{analysis state} which records the current state of the visualization along with the \rev{annotations made by the user in the form of comments. These comments serve to capture, for example, the user's assumptions and ideas or tasks to be performed. Users can also annotate the investigation progress by creating plain comments (i.e., unrelated to visual analysis) serving as logs, e.g., to mark down when a request for certain data was made.
Such externalization process proved to be of critical importance especially in the collaborative analysis~\cite{Mahyar:2012,Mahyar:2014}}.

The Progress tracking diagram then provides an overview of all captured states \rev{and comments}, organized into individual branches that reflect the natural divergence of user's thinking and multi-user collaboration. These can be browsed and revisited by users at any point in the investigation to get an overview or to be further analyzed.
The progress tracking supports two basic modes of the exploratory process, denoted as the divergent and convergent modes. 
In the divergent mode, criminalists search for possible scenarios, formulate hypotheses, and try to get various perspectives on data that can contain valuable insights. This means they create new branches reflecting various paths of exploration with many analysis states, capturing different views on data. \rev{When a branch represents a single hypothesis, the branch's description is then used for the hypothesis formulation and can be revisited at any point.}
In the convergent mode, criminalists create inferences, confirm or reject hypotheses, and combine individual paths of exploration to assemble evidence. This naturally corresponds to the merging of individual branches.

The progress tracking also supports the non-intrusive collaboration mode. Criminalists can share their work with colleagues who can see their progress diagrams and browse through analysis states in the read-only mode. It is intentional that only the author of the analysis state can edit it as it is usually tightly connected to the investigator's mental process. However, if criminalists want to continue the analysis on a colleague's state, they can merge it into their branch.
Last but not least, as new evidence and findings are added to the investigation, the progress tracking also informs users about the updates and changes of data related to the case, which may influence the ongoing investigation and visually indicate potentially affected analysis states.

\subsection{Data Credibility Levels}
\label{sec:credibility}
\rev{In the process of building evidence, it is pivotal that criminalists are well informed}
about the source of the data and how much they can trust the source \rev{(requirement R3)}. Based on that, they can invoke further steps to increase credibility, such as a search for additional data or witnesses.

The primary source of data is the central database that stores all data captured and related to a given case. 
Criminalists are commonly using the evaluation system based on the 4x4 matrix~\cite{matrix} \rev{evaluating the credibility of the source of evidence and of the information itself.}
This matrix gives investigators valuable information about the trustworthiness of data that is stored in the central database. 
However, criminalists also need to face other data-related problems, such as data inconsistency, incompleteness, and incorrectness, described in Section~\ref{sec:background}. To deal with these, they need ways to insert, modify and annotate data within the investigation. However, as the central database contains only verified information and cannot bear temporarily added and unproven information, criminalists need means to insert data that is in a state of an assumption or with insufficient proof directly into the tool. Such data can represent whole objects, such as a person or real estate, or can be in the form of individual attributes of already existing data objects, for example, the tax number of a person stored in the central database. Naturally, it is \rev{vital} that such inserted data needs to be strictly distinguishable from the data taken from the central database. 

\begin{figure}[tb]
  \centering
    \includegraphics[width=0.6\linewidth]{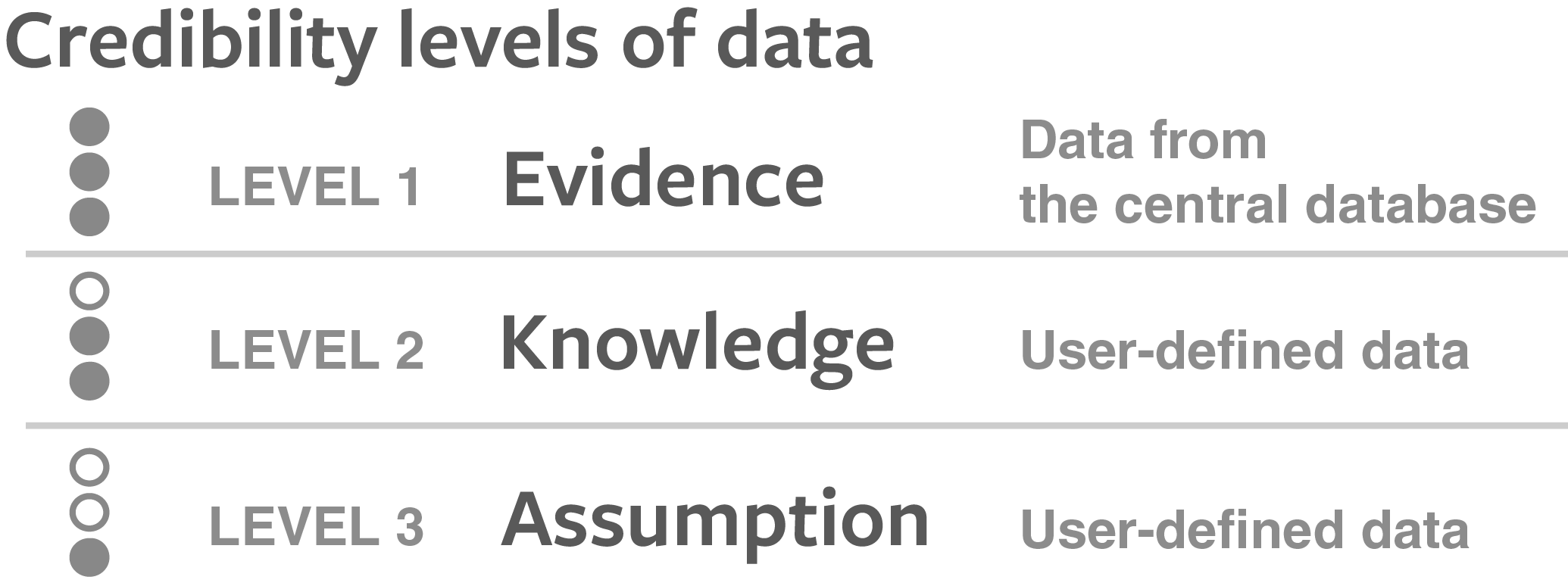}  
\caption{Three levels of data credibility, visually encoded by three signaling dots. The filling of these dots represents the levels -- the number of filled dots represents the credibility level: the more dots, the higher credibility. Such visual notation is used throughout the Visilant tool (see Section~\ref{sec:visilant}).}
  \label{fig:credibility}
\end{figure}

Furthermore, we can distinguish between two types of data inserted by investigators.
The first type corresponds to data added when criminalists introduce their \textit{assumptions} or when they want to test how the presence of certain data might affect the currently studied network. In this case, the data has a low or no evidence value and therefore should not affect other analysis states or collaborators' work unless requested on demand.
The second type of user-introduced data represents information that is either well-known or already trusted by criminalists, however, it is not yet formally proven, or the evidence has not been fully acquired. We denote such data as \textit{knowledge}, an unofficial term used by our expert collaborators.
Based on these types and discussions with criminalists, we introduce three basic \textit{credibility levels} of data (see Figure~\ref{fig:credibility}), which are treated differently in the system:

\vspace{1mm}
    \noindent\textbf{Level 1: Evidence   } Data stored in the central database with the highest credibility. It is available to all analysis states, branches, and criminalists.

    \vspace{1mm}
    
    \noindent\textbf{Level 2: Knowledge   } Trusted information among the team of criminalists, which occurred within the investigation and it is not coming from the central database. The trustworthiness of this information is already so high \rev{that it is desirable to share it throughout }all branches and with colleagues. It needs to be visually distinct from the visual representation of evidence.
    
    \vspace{1mm}
    
    \noindent\textbf{Level 3: Assumptions   } Data that exists only temporarily or is poorly supported by evidence. Its visibility is therefore limited only to the user who inserted it and it exists only in the analysis state where it was introduced and in the states derived from this one. It needs to be visually distinguishable from the visual representation of evidence and knowledge.
    
    \vspace{1mm}
    
To clearly visually distinguish between these three types, we encode them with three signaling dots (see Figure~\ref{fig:credibility}).
As the level of trustworthiness in individual data entries can change, users need to have the option to change the level of credibility with respect to official operational processes. Therefore, inside the tool, users can only either promote the assumption to knowledge on-demand, which is usually followed by filling in a source of information, or demote the knowledge back to assumption when necessary. Whenever this happens, the other investigators need to be notified about it so they can review the changes and incorporate them into their own visual and mental representations. 
A promotion to evidence (level 1) requires launching the process of official acceptance of the data into the central database.

\rev{\section{Framework for Criminal Analysis}}
\label{sec:framework}
\rev{Before describing our Visilant tool in detail, we need to introduce the whole framework supporting criminal investigations that was designed and tested along with Visilant. 
In this chapter, we will briefly introduce the main components and functions of this framework to give a better picture of the overall capabilities of our solution.

The framework aims to provide the criminalists with a unified analysis of large-scale heterogeneous data, centralized around the network representation connected to individual analytical modules. These modules represent a variety of analytical tools divided them into two groups -- automatic and interactive ones. Automatic modules include, for instance, the automatic extraction of photos from storage media, indexation and classification of videos, and correction of corrupted traffic traces. The automatic modules prepare the data for interactive modules that already require an appropriate visual representation, provided by Visilant. These can, for example, identify groups of people communicating via phone calls or performing financial transactions. All modules are coordinated via an orchestrator, ensuring efficient use of modules and their calls. As a whole, the framework integrates the most commonly performed operations on the captured data that were revealed within the requirement analysis.
More technical details about the framework are to be found in the supplementary material.

As already stated, the central part of the framework is the network visualization, enabling to explore relationships between individual objects. However, some objects possess additional information, such as the timestamp or GPS location, thus, the tool should provide the criminalists with additional views exploring these phenomena. The analytical part of the exploration of these types of data is already available in our framework and their visual representation is currently in progress, as this was stated as a very desirable extension of Visilant within the conducted online evaluation, summarized in Chapter~\ref{sec:feedback}. 
}

\vspace{2mm}
\section{Visilant Tool}
\label{sec:visilant}
\begin{figure}[tb]
  \centering
    \includegraphics[width=1.0\linewidth]{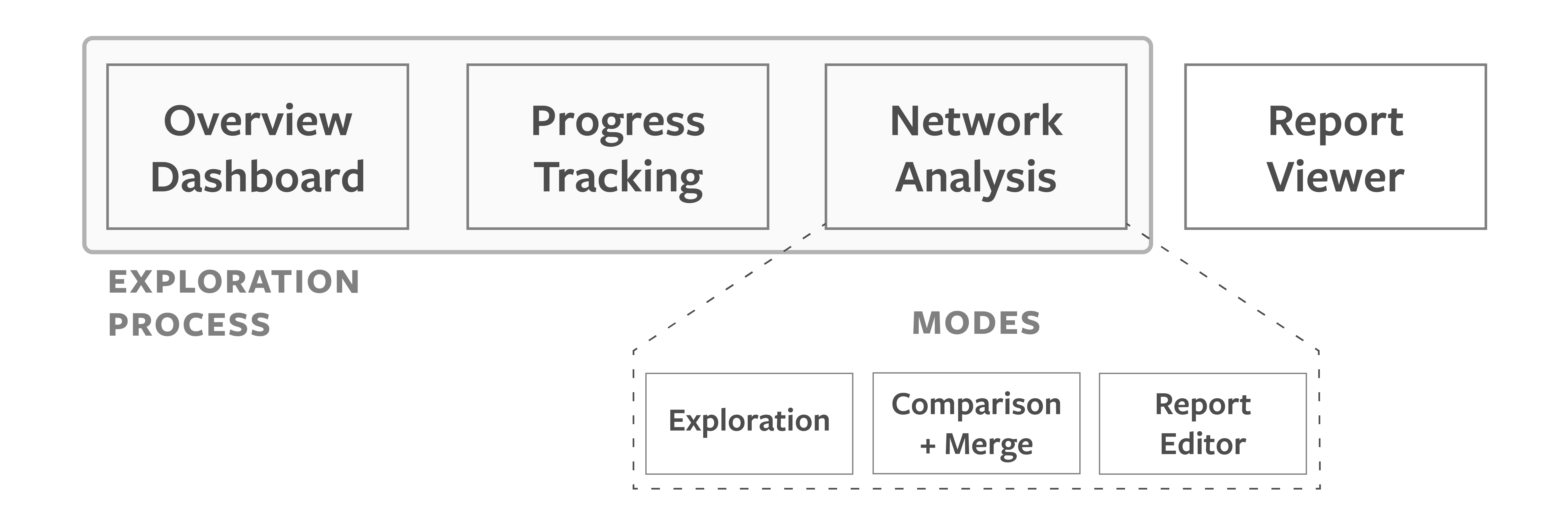}  
\caption{Overview of the Visilant components. The exploration is supported by the Overview dashboard, Progress tracking, and Network analysis. At the end of the investigation, the user can generate and view the final report in the Report viewer, and then export it for state attorneys.}
  \label{fig:overview}
\end{figure}

Based on the requirements stated in Chapter~\ref{sec:requirements} and our proposed design concept, we implemented a web-based Visilant tool, providing the interactive visual support of the criminal investigations.
\rev{The tool is composed of a front-end, implemented using the framework Vue.js and its core libraries, and a minimalist back-end. On the back-end, the tool fetches information about available data, visualization documents, and reports from the framework's data storage along with currently analyzed data in the form of objects and relationships. 
The graphical user interface, as well as the whole network visualization, is built using the features of the front-end framework, except for the force-based calculation of the graph layout which uses d3-force.} To support the given tasks, our tool is composed of three main views, the Overview dashboard, Progress tracking, and Network analysis, accompanied by the Report viewer (see Figure~\ref{fig:overview}). 

The Overview dashboard (Section~\ref{sec:dashboard}) serves as a landing page for the whole investigation of a given case. The Network analysis view aims for the visual exploration of data on the lowest level, providing users with three working modes, described in detail in Section~\ref{sec:network}. 
These views are accompanied by the Progress tracking diagram (Section~\ref{sec:tracking}) enabling to trace the investigation scenarios, their progress and branching, and to revisit an arbitrary analysis state stored within the investigation. 

\subsection{Overview Dashboard}
\label{sec:dashboard}
\begin{figure*}[htb]
  \centering
    \includegraphics[width=1.0\linewidth]{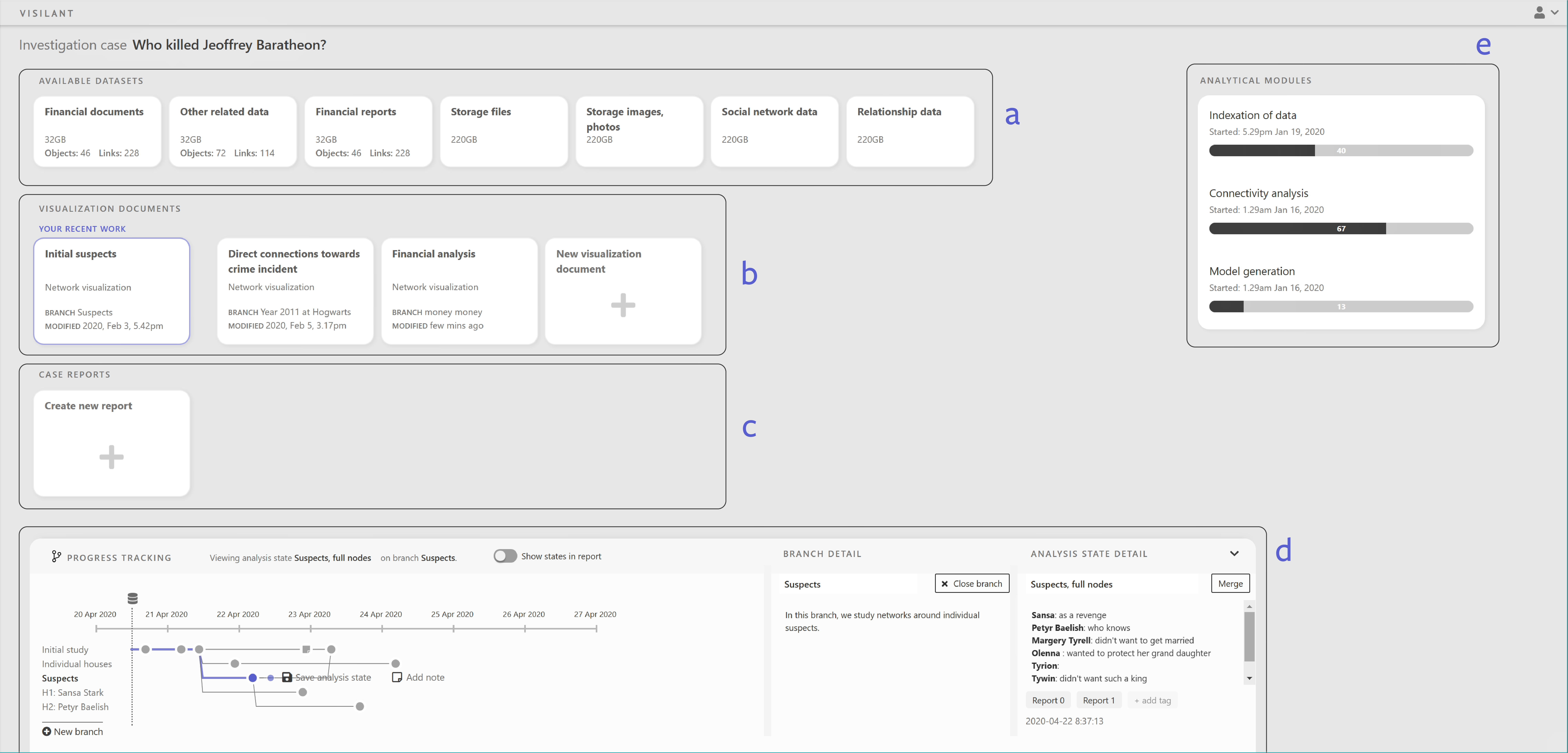}  
\caption{Overview dashboard containing a) a list of available datasets in the central database, b) a list of already created visualization documents and the option to create a new one, c) a list of already generated reports and the option to create a new one, d) the access to the Progress tracking diagram \rev{along with the active branch and commit description}, and e) a list of currently running data analysis jobs at the back-end, preparing the data for the exploration process.}
  \label{fig:dashboard}
\end{figure*}

Overview dashboard (Figure~\ref{fig:dashboard}) represents the initial entry point to the whole exploration process of Visilant. Here the user gets the overview of all datasets that are available for the given case in the central database (Figure~\ref{fig:dashboard} a). 
Then, the user can create a new visualization document, i.e., to start a new exploration process on data from the central database. 
\rev{Both the data queries and available data analyses, e.g., community search, are performed on the back-end}
part of the whole platform, where Visilant serves as the front-end interface. 
The list of running jobs at the back-end and their status is displayed in the right part of the dashboard (Figure~\ref{fig:dashboard} e). 

The Overview dashboard then lists available visualization documents that the user can select to further work on (Figure~\ref{fig:dashboard} b).
From the dashboard, users can also generate final reports for state attorneys or other third parties and view the already created ones (Figure~\ref{fig:dashboard} c). 
The last part of the dashboard landing page is formed by the Progress tracking diagram (Figure~\ref{fig:dashboard} d) that serves here as the first-glance view onto the status of the currently selected (active) visualization document.


\subsection{Network Analysis View}
\label{sec:network}
\begin{figure}[tb]
  \centering
    \includegraphics[width=1.0\linewidth]{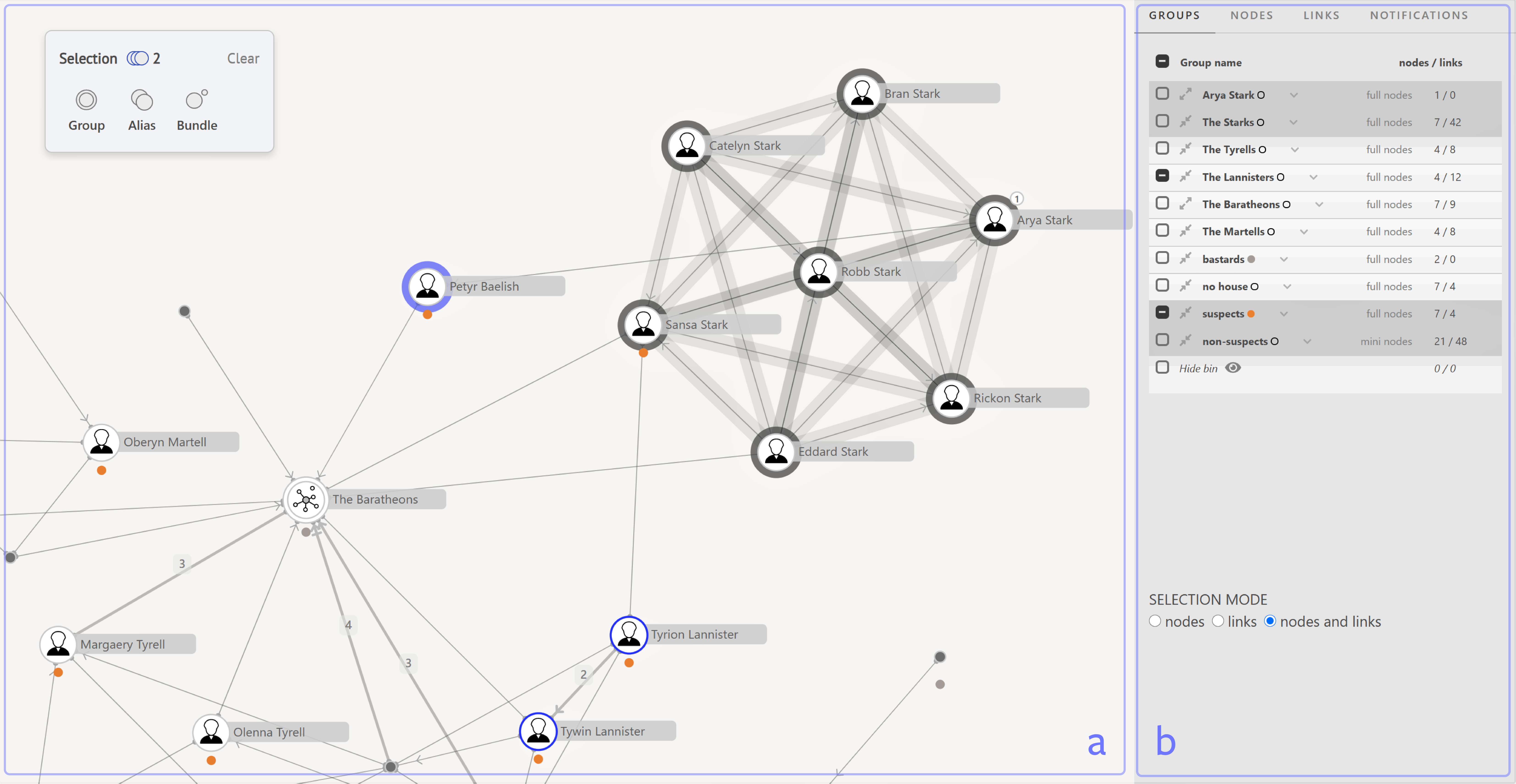}  
\caption{ \rev{Exploration mode of the Network analysis view, consisting of a) the central layout and b) the side panel containing additional information about selected objects.}\rev{ Nodes belonging to the current selection are outlined in a thin blue ring. Nodes representing key objects (focus nodes) are outlined with a thick blue ring for better visibility. Nodes which are currently not relevant for analysis can be minimized into small dots.}}
  \label{fig:network}
\end{figure}

\begin{figure*}[htb]
  \centering
    \includegraphics[width=1.0\linewidth]{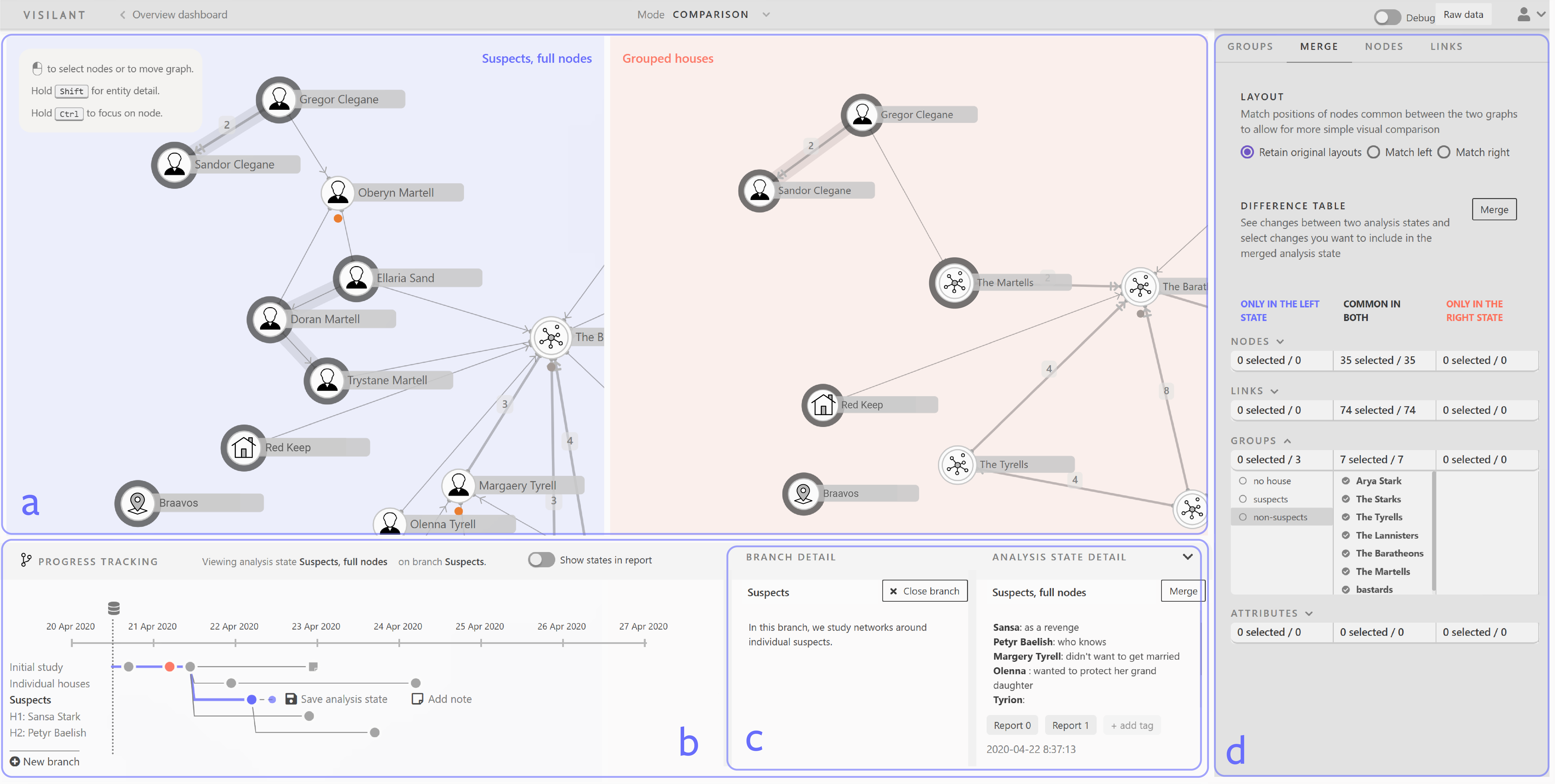}  
\caption{\rev{a) Network analysis view in the comparison mode for exploring similarities and differences between two selected analysis states.}\rev{ b) The Progress tracking diagram displays the currently viewed analysis states in orange and blue and c) active branch and analysis state details. d) The right-side panel shows details about the comparison of the two states and provides means for merging these to a single network.}}
  \label{fig:comparison}
\end{figure*}

Network analysis view (Figures~\ref{fig:network} a, \ref{fig:comparison} a) is a visual component that, along with the network visualization, is the key technique used in criminal investigations. It provides users with a set of specifically designed visual and interactive features that support the exploration process. 
When designing its visual appearance \rev{(Figure~\ref{fig:network} a)}, we followed the principle of minimalism, avoiding excessive use of colors when not necessary, utilizing simplified icons, and keeping the layout as clean as possible.
\rev{This decision was based on the study of the network visualizations currently used by criminalists and existing studies and guidelines~\cite{Lee:2013,DrelieGelasca}.}
With the increasing number of nodes and links, the representation can become very quickly cluttered, and it then makes it more difficult to achieve fine-grained interactions on individual objects, such as highlighting or focus+context. Therefore, all additional information about the nodes and links can be viewed on-demand, either by hovering or clicking on a given node or link or by viewing it in the right-side panel \rev{(Figure~\ref{fig:network} b)}.

\rev{On the first load of data into the network visualization, we use the force-based algorithm for the initial layout of nodes. Throughout the analysis, the layout is frozen and can be changed only on users' demand as 
unprompted changes of layout would significantly influence the perception of the relationships. It is thus} essential for the work of criminalists that their network analysis does not change without their intention. This significantly helps to reduce the extraneous load and saves their mental effort for more difficult tasks.

Except for the common interaction options, our network representation supports several advanced operations to fulfill the requirements \rev{R7, R8, and R9}, stated in Chapter~\ref{sec:requirements}. These include the grouping of nodes to reduce the complexity of the network appearance \rev{and to decrease the visual clutter} by introducing logical collections of data \rev{(requirement R9)}.
The grouped node then contains the annotation aside with the number indicating the amount of aggregated nodes. The content of the grouped node can be expanded when necessary or explored in the right-side panel, providing users with the supplementary information about all objects in the network visualization. Moreover, users can assign individual groups a tag color which is then displayed below the node in the visualization. This can be helpful when users study the network visualization and can be easily visually informed that the node belongs to a certain group, e.g., a group of victims or suspects. Moreover, the user can minimize the visualized nodes, which are then represented by small dots \rev{(Figure~\ref{fig:network} a)}. This allows to temporarily reduce currently unnecessary information in the network and to bring focus on the analyzed data, while still keeping the original structure of the network.
Users can also visually highlight key objects in the network so they can be easily spotted among other nodes. These are then displayed with a bold color outline around the nodes \rev{(see Figure~\ref{fig:network} a)}.

User can also introduce new objects into the network \rev{(requirements R7 and R8)}. Keeping in mind the data credibility levels, these need to be clearly visually distinguishable from the data coming from the central database. In agreement with the experts, we decided to use the dashed outlines to indicate user-defined objects, as that is the convention in the tools they commonly use. The Network analysis view offers three modes: \begin{itemize}
    \item \textbf{Exploration mode} enables users to study and interact with single network visualization. It supports all the above-mentioned interactions.
    \item \textbf{Comparison mode} allows users to display two network visualizations side by side and enables their comparison and user-driven merging  \rev{(requirements R2 and R5)}. As this is one of the core functionalities of Visilant, it is described in detail in the remainder of this section.
    \item \textbf{Report  mode} \rev{(requirement R6)} provides a way to create a final report for external authorities from selected analysis states. It enables to append textual descriptions accompanying the \rev{analytical states} that appear in the final report as a sequence of individual views on data.
\end{itemize} 

At a point when it becomes \rev{necessary} to see two analysis states at the same time, e.g., when comparing results of similar hypotheses, merging the findings with colleague's results, \rev{or integrating findings from old cases}, the investigator switches into the comparison mode to explore the similarities and differences between the corresponding network representations (see Figure~\ref{fig:comparison}). 
The right-side panel provides two copies of the same views as in the exploration mode, \rev{i.e., the group, node, and link panels,} one for each network. Moreover, to aid the comparison, we provide users with an additional \rev{Merge} panel showing the list of commonalities and differences \rev{in the Difference table} (see Figure~\ref{fig:comparison} \rev{d)}. 
\rev{The Difference table shows lists, divided into three parts, presenting similarities and differences between the two states. The middle part of the list shows objects that are equal in both states (based on the sets of attributes and their values), and then, for each state, there is a part that shows}
state's unique objects. 
The observation is further supported by linked highlighting in all views.
In this stage, the user can either decide to return to the exploration mode or to finalize the merge operation, which results in a new analysis state. For the merge operation itself, the user selects which of the distinct parts of the two networks should be included in the newly merged state \rev{in the Difference table}. For example, when the two analysis states' nodes are organized into different groups, users can select which of these groups will propagate to the merged state -- they can include all, just some, or none. Also, while they may include groups just from state A, they may decide to include the nodes only present in state B and discard any extra nodes from state A. Users also select which one of the two original network layouts will be adopted for the merged state.

\subsection{Progress Tracking Diagram}
\label{sec:tracking}
\begin{figure}[htb]
  \centering
    \includegraphics[width=1.0\linewidth]{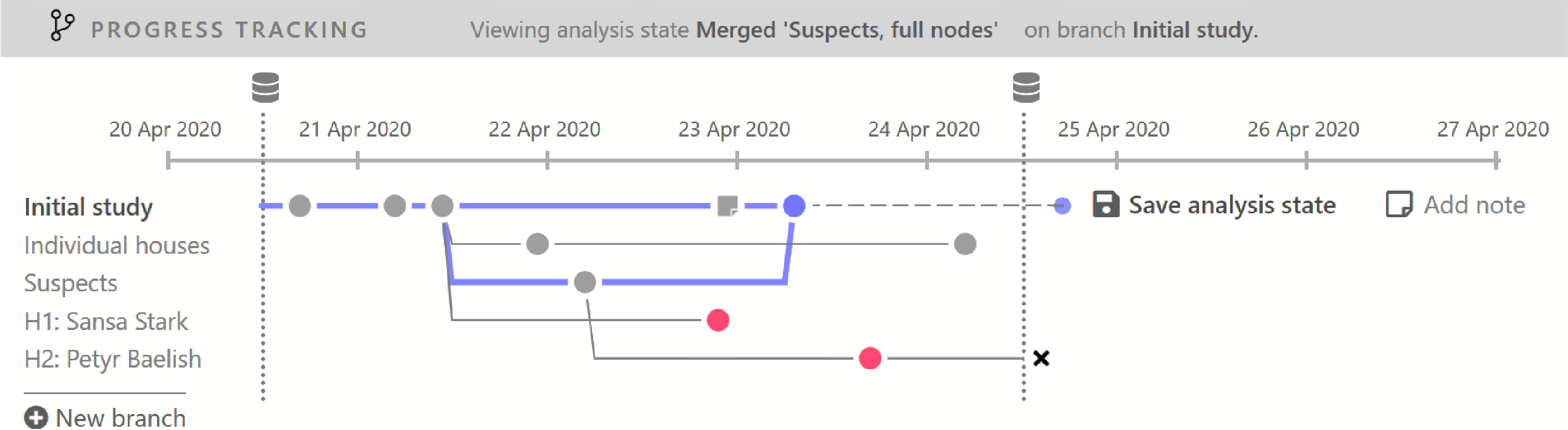}  
\caption{Progress tracking diagram showing an overview of the current investigation state. The active state and active branch are highlighted in blue, states highlighted in pink indicate that some of the data contained in the view has been updated and needs to be explored by the user.}
  \label{fig:tracking}
\end{figure}

The Progress tracking diagram (Figure~\ref{fig:tracking})\rev{, supporting requirement R1,} is specifically designed to keep the big picture of the investigation progress, its current state, and its history. It is displayed at the bottom part of the tool and can be collapsed
when not actively used to make more space for the analysis of the network.
The diagram depicts individual branches of the investigation where each branch may contain any number of analysis states represented by nodes in the diagram.
Each node bears information about the state of the network visualization, including all operations applied, such as groupings, or annotations. Therefore, by simply selecting these individual nodes, the user can jump back to the given state of the network and further interact with it. 

The diagram is arranged in line with the timeline, so the temporal continuity of the states is clearly visible.
The users can create a new branch whenever they want to start a new path of exploration. The branch name is displayed on the left side of the diagram, aligned with the branch path.
A new branch can start from the initial node corresponding to the initial investigation point of the current visualization document or it can be branched from any existing analysis state. The latter option is very common, as in different stages of the investigation, several possible scenarios can appear and need to be explored.
The active branch is highlighted in the diagram and users can expand it with new analysis states derived from the last ones or just create a note containing additional information related to the exploration, which is also displayed on the branch path. For both operations, users can find buttons at the end of the branch path in the diagram.
The Progress tracking diagram also displays details on the active branch and its active state \rev{(Figure~\ref{fig:comparison}~c)}. These can be further edited by the user.

By selecting two nodes of the diagram simultaneously, the user is automatically directed to the comparison mode in the Network analysis view, where the two states can be compared and potentially merged. The merge operation is then clearly visible in the Progress tracking diagram, as the node representing the merged state has two branches entering it (Figure~\ref{fig:tracking}). 

The Progress tracking diagram also indicates updates of analysis states and their deletion, since the data is likely to change during the investigation, e.g., when new information is loaded into the central database \rev{(requirement R4)}.
Also, at any point during the exploration, the investigator can select any analysis state to be included in the report and then switch to the report editor mode of the Network analysis view to append the description for final report generation.


\section{Evaluation and Expert Feedback}
\label{sec:feedback}
\rev{The usability of our proposed Visilant tool was evaluated within two qualitative studies conducted with two experts in criminology -- the head of the criminal intelligence analysis in the national law enforcement agency of the NCCOC of the Czech Republic and his deputy. 
The head criminalist was also heavily involved in gathering the initial requirements and we repeatedly consulted with him the design and progress of Visilant development. 
On the other hand, his deputy was introduced to the Visilant tool only at the evaluation session.
The first evaluation was conducted online due to the COVID-19 outbreak.} 
\rev{The next round of evaluation was in-person with the above-mentioned experts. 
Within this session, we revealed additional interesting observations.}

\subsection{First Round -- Online Session}
During the online evaluation, we presented the experts the tool and its functions and \rev{asked} for both benefits of the tool, as well as its drawbacks and suggestions for further improvements.
\rev{As the criminological data is highly confidential, we had to conduct the study on a testing dataset commonly used within the whole project. }

The experts highly appreciated that Visilant targets the bottlenecks in their daily workflow and they expressed the desire to test Visilant on a variety of their cases. 
They especially appreciated the possibility to track the history of the investigation by creating branches in the Progress tracking diagram and saving the important analysis states \rev{(R1)}. This is a highly useful feature, 
as it is crucial to formulate several hypotheses at the beginning of the investigation rather than to follow the seemingly obvious one, as it can often turn out to be a blind alley. The experts confirmed that the creation of a set of possible scenarios is a very common approach and they appreciated that we support that. 

In the exploration mode of the Network analysis view, they stressed the necessity to distinguish between the credibility levels \rev{(R3)} visually, discussed in Section~\ref{sec:credibility}. They confirmed that the three-level credibility design exactly reflects their needs and agreed that the current visual depiction is appropriate and, in combination with the information from the 4x4 matrix, it provides further support for their investigation. 
To support the investigation process even more, they suggested that it would be extremely useful to also consider data-based layouts for networks, such as geographical or temporal. Projecting the objects to the corresponding positions on the map layout and connecting it with the presence of the object at a particular time would help to reveal missing data or very interesting events, otherwise difficult to notice. \rev{As the criminalists are well accustomed to the concept of space-time cube~\cite{Kraak:2005}, we plan to integrate this concept into our framework as well.} Another valuable extension would be the option to operate with a set of photographs, attached to a particular object as a thumbnail of the node and fully displayed when the user zooms in. 

\rev{When evaluating the interactions within the Network analysis view in its exploration mode, they highly appreciated the option to enter new pieces of knowledge to the view (R7) and the visual distinction of newly added nodes and links (R4). This helps them to clearly distinguish between the data coming from the database and assumptions made by the criminalist within the investigation. 
Also, the option to create a placeholder object (R8) was appreciated as it is common that they need to add an object with no clear identifier but whose connections to other objects are already known.
They also commented on the possibility to semantically group objects (R9) that enables them to reduce the visual clutter and focus on the crucial aspects and objects.}
The comparison mode of the Network analysis view and its merge operation \rev{(R2)} was confirmed to be substantial for the support of individual stages and progress of the investigation, sharing the evidence and knowledge among colleagues, and compiling the final report for state attorneys. They found the visual highlighting of similar and distinct objects between the two networks very helpful and they stated that it would be beneficial to have advanced options for changing one network’s layout based on the other one. When discussing the comparison of two networks, the experts immediately started discussing options to investigate the differences among more than two network states at once. After vivid discussion, they concluded that the current solution, enabling to compare and merge two networks, is the right solution, because of the potential cognitive load of the "multi-network" comparison.
\rev{The criminalists valued that Visilant strictly follows the requirement R5 as their investigation process cannot be influenced automatically by adding new information whenever it is available. 
This has to be invoked only on demand, as implemented in Visilant.}
They also appreciated the option to compile and export the report from the investigation \rev{(R6)}.
This will significantly ease the important and tedious parts of their workflow and free up more time for the investigation itself.

\subsection{\rev{Second Round -- On-site Session}}
\rev{Three months after the online session, we managed to arrange a personal meeting with the same investigators participating in the first session. 
This time, we had a chance to directly present the Visilant tool to them. Although the testing was still conducted on a synthetic dataset, the personal evaluation gave us additional insights. 
Except for confirmation of the first evaluation results, the second session revealed several new interesting aspects.
Testing the Progress tracking diagram identified a new need for handling hypotheses. The criminalists would appreciate the possibility to accept or reject a hypothesis only partially and see the corresponding visual feedback in the diagram.
The comparison mode of the Network analysis view was highly appreciated as the possibility to overview the relationships in both views and seeing the commonalities and differences in the side view are necessary for subsequent decisions about objects to be merged.
Testing the merge option of the comparison mode revealed an additional request for collaboration. Except for loading only a single analysis state of the investigation that could be merged to the local one (as implemented in the current version of Visilant), they would appreciate the option to load the whole visualization document of a colleague or from past cases with its history (i.e., its Progress tracking diagram). The criminalist would thus gain context for a better understanding of the given analysis state.

Within this round, the experts confirmed that the tool is useful for both short- and long-term investigations. 
Even when the short-term case is obvious and only a few entities are involved, it is still worth to generate and store the Network analysis view of the case. 
This can pay off in future investigations when the criminalists check the criminal history of suspicious persons and objects.
The criminalists are building a complex knowledge base spanning throughout years, and new investigations trigger vast searches in this base and search for matches between objects.
The Visilant concept is again in line with this as it visually supports the search and merge tasks.}

\section{Conclusion}
In this paper, we addressed the problems and challenges related to the domain of criminal investigations in the context of the visual aids that can be offered by the visual analytics approaches. 
We derived \rev{key} requirements and the design concept for the creation of visualizations supporting them. The concept was followed in the design of Visilant, which offers several linked views supporting the exploration process of the investigations. The \rev{two evaluations}, conducted with the chief criminalists, proved that the design concept, as well as its projection to the Visilant tool, \rev{supports their daily workflow in the initially envisioned way}. In combination with the back-end data analyses \rev{performed within the whole framework, it can become} a very powerful and robust assistant for their future investigations.



\acknowledgments{
The presented work has been supported by the Ministry of Interior of the Czech Republic under the research project no. VI20172020096.}

\bibliographystyle{abbrv-doi}

\bibliography{vislant}
\end{document}